\documentclass[%
 reprint,
amsmath,
amssymb,
aps,
floatfix,
superscriptaddress,
]{revtex4-2}
\usepackage{graphics}
\usepackage{graphicx}% Include figure files
\usepackage{dcolumn}% Align table columns on decimal point
\usepackage{bm}% bold math
\usepackage{mathrsfs}
\usepackage{pstricks}
\usepackage{color}
\usepackage{orcidlink}
\usepackage{slashed}
\usepackage{amsmath}
\usepackage{epsfig}
\usepackage{amsfonts}
\usepackage{amssymb}
\def\beq{\begin{equation}}
\def\eeq{\end{equation}}
\def\bea{\begin{eqnarray}}
\def\eea{\end{eqnarray}}

\RequirePackage[normalem]{ulem}
\RequirePackage{color}

\begin{document}

\title{Hydrodynamic fields in fluctuating environment:\\ a model for isochoric heat capacity of simple liquids}

\author{I. P. de Freitas
\orcidlink{0009-0007-2532-2929}}
\email{Email address: isaquepfreitas@cbpf.br}
\affiliation{%
Centro Brasileiro de Pesquisas F\'{\i}sicas, 22290-180 Rio de Janeiro, RJ, Brazil}

\author{F. Sobrero
\orcidlink{0009-0009-6497-6612}}
\email{Email address: felipesobrero@cbpf.br}
\affiliation{%
Centro Brasileiro de Pesquisas F\'{\i}sicas, 22290-180 Rio de Janeiro, RJ, Brazil}

\author{A. M. S. Macedo
\orcidlink{0000-0002-4522-031X}}
\email{Email address: antonio.smacedo@ufpe.br}
\affiliation{%
Universidade Federal de Pernambuco - UFPE, Recife, PE, Brazil}

\author{N.~F.~Svaiter
\orcidlink{0000-0001-8830-6825}}
\email{Email address: nfuxsvai@cbpf.br}
\affiliation{%
Centro Brasileiro de Pesquisas F\'{\i}sicas, 22290-180 Rio de Janeiro, RJ, Brazil}
%%

%%%%%%%%%%%%%%%%%%%%%%%%%%%%%%%%%%%%%%%%

\begin{abstract}

Using functional methods, we investigate the sound quanta arising from quantized hydrodynamic fields in simple liquids at low temperatures, under the influence of high-energy processes, coming from non-hydrodynamic degrees of freedom. To model these effects on the hydrodynamic fields, we assume that the quantum fields are coupled to an additive, delta-correlated (in space and time) quantum noise field. Thus, the hydrodynamic fields are defined in a fluctuating environment. After defining the generating functional of connected correlation functions in the presence of the noise field, we perform a functional integral over all noise field configurations. We obtain a new generating functional written in terms of an analytically tractable functional series.
This formalism allow us to obtain the excitation spectra of liquids. In the liquid, each term of the series describes the emergent non-interacting elementary excitations with the usual phonon-like dispersion relation and additional excitations with dispersion relations with gaps in pseudo-momentum space. Finally, the behavior of the constant volume specific heat as a function of temperature is obtained for different simple liquids.

\end{abstract}

%%%%%%%%%%%%%%%%%%%%%%%%%%%%%%%%%%%%%%%%


\maketitle

%%%%%%%%%%%%%%%%%%%%%%%%%%%%%%%%%%%%%%%%

%\section{Introduction}

\section{Introduction}
\label{introduction}\label{intro}

 Liquids are strongly interacting, dynamically disordered systems. They exhibit different heat capacities as a function of temperature. Their different structures make it quite difficult to adopt a general model for the constant volume heat capacity. Here we show that the connection between phonons and thermodynamic properties describing solids can also be applied to liquids. By discussing the contributions coming from quasi-particles with gaps in pseudo-momentum space and phononic excitations, the temperature dependence of the isochoric specific heat is presented for liquids with simple structure. 
 
 Using the functional integral formalism of field theory \cite{matheus,yaglom,fradkin,martin,livro5,roepstorff,zinnjustin},
we establish, in low-temperature liquids, a connection between gapped momentum states of elementary excitations and hydrodynamics via an effective field theory \cite{baggioli2,bra3}. To model the effects of high-energy processes coming from non-hydrodynamic degrees of freedom, we assume that the  hydrodynamic fields are coupled to an additive delta-correlated noise field. 
In the presence of the noise, we define an 
augmented generating functional of connected correlation functions.  Performing the functional integral over all configuration space of the noise field in this augmented generating functional, we obtain 
a functional series representing a new generating functional. In the functional series, we characterize effective actions describing emergent phononic excitations with the usual dispersion relation (i.e., sound quanta) and collective excitations with dispersion relations exhibiting gaps in pseudo-momentum space. With our approach, performing a configurational averaging procedure, we describe the gapped momentum states discussed by other methods, such as, for example, the Keldysh-Schwinger approach to dissipation  \cite{calzeta,kamenev}. 

Contrary to the situation of crystals where the vibrational motion is decomposed into independent normal modes, due to a unique length scale, in liquids
there is no small parameter to implement a perturbative expansion \cite{lan1}. 
In liquids, the displacements are large, and the interatomic/intermolecular interactions are strong, comparable to those of gases. To find the temperature dependence of the thermodynamic quantities in liquids, one cannot expand the potential energy of the liquid in terms of the square of atomic/molecular displacements. To make things more complex, liquids are characterized by diffusive phenomena on short time scales and exhibit viscoelastic properties. A system with features of both viscous fluids (that generate shear stress for an inhomogeneous flow velocity) and an elastic body (that produces a shear stress in a static state).  Therefore, for a sufficiently short time scale, a liquid may be modelled as an amorphous solid with disorder. One common approach to describe a disordered solid is to use random differential equations \cite{stephen,gu,blaunstein,altland,cheng,conf}. Randomness can also be used in liquids, since, as we discussed, they are strongly interacting, dynamically disordered systems \cite{strong}. Their behavior is similar to that of an amorphous solid without the emergence of solidity. 
These considerations led us to study the excitation spectra at short wavelengths of low-temperature liquids by introducing randomness to model a fluctuating environment. We introduce noise not in the differential operator but as an additive, delta-correlated noise field. 

The conventional approach to studying classical liquids is based on the general formalism of time-dependent correlation functions and linear response theory \cite{hansen,ortiz}. In the framework of linearised hydrodynamics one can obtain the macroscopic transport coefficients in terms of the microscopic quantities.
Defining $\tau_{c}$ as the mean collision time of the constituents of the liquid and the wave number-frequency $\omega$, for $\omega\tau_{c}\ll 1$ one uses an effective field theory, bounded from above by some energy scale, with the hydrodynamic fields defining a continuum classical field theory \cite{lan2,vo,hy}. 
One approach to use the hydrodynamic fields in the regime $\omega\tau_{c}\geq 1$, on the molecular scale, is the generalized hydrodynamics, which considers frequency and wave number-dependent transport coefficients. Upon leaving the regime where liquids are not able to oppose tangential stresses, it is possible to show the appearance of shear waves in liquids \cite{torell1,torell2}.
Another way to access the regime $\omega\tau_{c}\geq 1$ was presented by Frenkel \cite{frenkel,frenkel2}, with the propagation of solid-like collective modes in liquids, using Maxwell's analysis. Maxwell discussed a  simple model for viscoelastic materials that exhibit a behavior between a purely viscous liquid and an elastic solid. In liquids, there is a viscous flow on long time scales and elastic behavior on very short time scales. Frenkel defines $\tau_{f}$, the liquid relaxation time, i.e., the average time that atoms/molecules spend traversing  the interatomic/intermolecular spacing. For times shorter than $\tau_{f}$, the behavior of the system is that of a disordered solid with a rigid disordered structure, with shear elastic waves. There is an interpolation between the purely elastic solid behavior and the purely dissipative response of a fluid. 
Using this approach, it is possible to obtain a microscopic picture of the liquid state with a dispersion relation exhibiting gaps in momentum space, which for instance has been discussed in different areas of physics \cite{aharonov,ben,chiao,bra,kt,yang,baggiori,experiment,naimark}. 
The energy spectra of such systems share some similarities with the tachyon spectrum in quantum field theory \cite{t1,meta,t2,t3,t4,t5,re0,t8,t10}. 
At this point two interesting questions may be formulated: (i) can we include in the model the effects of degrees of freedom associated with the underlying microscopic theory without making use of generalized hydrodynamics? (ii) Still using hydrodynamic fields, is it possible to obtain the gapped momentum states without using the Maxwell-Frenkel viscoelastic theory?

Using the functional formalism, we develop a theoretical framework to
obtain the bosonic excitation spectra of low-temperature simple liquids in the regime with frequencies satisfying $\omega\tau_{c}\geq 1$, i.e., the short-time behavior of the correlation functions. 
Our approach is an oversimplification 
of the effects of high-energy processes over sound quanta, when quantum effects are dominant.
To take into account 
short-time processes, we define an effective model of non-hydrodynamics degrees of freedom, introducing an additive noise field. This noise field represents unknown quantum processes at small distances or a quantum vacuum noise \cite{davies,noise}. 
Using the definition of the usual generating functional of connected correlation functions, one defines a generating functional in the presence of the noise field, an augmented generating functional. 
After integrating out the noise, we obtain a new generating functional, written in terms of a functional series. In each term of the series, one can show that there are two kinds of noise-induced quasi-particles. Those obeying the usual phonon-like linear dispersion relation and also elementary excitations with dispersion relations with gaps in pseudo-momentum space respectively. Considering the low temperature regime, we discuss the role of the emergent phononic and excitations with gaps in the quasi-momentum space \cite{ZacconeSHL, trachenko23, trachenko2008}. We are able to obtain the isochoric specific heat of liquids with simple structure, as for example Ar, Na and N$_2$.

Therefore, taking into account propagation modes with wavelengths comparable to interatomic separation allows us to obtain the constant volume specific heat as a function of the temperature. Our results suggest that, unlike in solids where the phase space available to phonons is fixed, in liquids this phase space is variable and reduces with temperature—a fundamental distinction that is consistent with the long-standing problems identified by Landau, Lifshitz and Pitaevskii regarding the absence of a small parameter in liquid theory combined with  strong interatomic interactions and dynamical disorder \cite{lan}. We have found evidence that 
a phonon-based approach that has been used for solids can also be used for liquids \cite{bolma2, bolmatov, Liu1, Liu2}.

The structure of this work proceeds as follows. In Sec.\ref{sec:disodered} we will briefly outline the quantization of the acoustic waves in liquids. In Sec.\ref{sec:disoderedLG}, we discuss the phononic field with the effects of an additive noise field, defining the augmented generating functional of connected correlation functions. In Sec.\ref{sec:thermal mass} we integrate out the noise in this generating functional, using a formal object, named  configurational zeta-function. In Sec.\ref{sec:two} we discuss the two-point correlation functions of the model with the emergence of phononic and gapped momentum states. The calculation of the specific heat in liquids with such excitations is presented in Sec.\ref{sec: specific heat}.  Finally, conclusions are given in Sec.\ref{sec:conclusions}. In \ref{sec:casimir}, we study the canonical quantization of the quasi-particles with gaps in the pseudo-momentum space. In \ref{sec: zeta regularization}, we perform the analytic regularization of spectral zeta-functions defined by the functional determinants of the model. 
\iffalse
A brief conceptual summary of our work is the following: to find the isochoric specific heat of liquids, one must deal with excitations, i.e., quasiparticles, that arise in two different regimes: the regime in which the hydrodynamic approximation is valid, and the other one in which it is not valid anymore, i.e., for characteristic lengths which are near the granularity size of the medium. We start from the hydrodynamic fields  and obtain the classical picture of sound waves. After imposing a non-commutative algebra of operators, we obtain the phononic quantum field and its Hilbert space of quasi-particle excitations. To discuss the excitation spectra of liquids taking into account the high-energy process, we adopt the functional integral approach. We define the generating functional of correlation functions $Z(j)$ by introducing an external source $j(t, \mathbf{x})$. Next, we define the generating functional of connected correlated functions $W(j) = -i\log{Z(j)}$. To model the effects of non-hydrodynamic degrees of freedom, we introduce a noise field $h(t,\mathbf{x})$ linearly coupled to the phonon field. The average of the generating functional of connected correlation functions $\mathbb{Q}[W(j)]$ is defined, and we obtain its series representation. In the $k-$th term of the series, there are $k-1$ phonon fields and one field with gap in the pseudo-momentum space. Using zeta-regularized products, the temperature dependence of the isochoric specific heat is presented. \fi In this work, we use the units $\hbar=k_B=1$.

\section{The acoustic wave quantization in liquids}\label{sec:disodered}

In spite of considerable efforts, a unifying physical model of liquid structure and its thermodynamic properties is still under construction, due to the complexity of the liquid behavior at different scales \cite{rolinson,peluso1,gchen}. As we discussed, liquids have viscoelastic properties, since on short time scales their behavior resembles that of an amorphous solid with structural disorder, while on longer time scales they behave as a viscous fluid. To investigate dynamical variables using space-time correlation functions and understand the microscopic structure of a liquid at the molecular scale, one must compare wavelengths with both the mean free path $l_{c}$ and the mean collision time $\tau_{c}$ of the liquid components. There are three different regimes for the wave numbers and frequencies. The region $k l_{c}\gg 1 $, $\omega\tau_{c}\gg1$, represents the free-particle regime where the distances and times involved in the processes are quite short. The components of the liquid move independently of each other. The range of intermediate wave numbers and frequencies, known as the kinetic regime, has $k l_{c}\approx 1$, $\omega\tau_{c}\approx 1$. For such frequencies, the wavelength is about the same size as the mean free path, which violates the assumption that the hydrodynamic fields are defined in the continuum. Therefore, one has to take into account the molecular structure of the liquid, and the current treatment is based on the microscopic equations of motion of the elementary components. Finally, the hydrodynamic regime where $kl_{c}\ll 1$, $\omega\tau_{c}\ll 1$.
In this regime, the behavior of the liquid is described by phenomenological equations for the hydrodynamic fields: the temperature, the mass density and the local velocity of the liquid, i.e., $T(t,\mathbf{x})$, $\rho(t,\mathbf{x})$ and $\mathbf{v}(t,\mathbf{x})$. One way to proceed is to develop the method of fluctuating hydrodynamics where we have a set of stochastic differential equations for the fluctuating variables $\delta\rho(t,\mathbf{x})$, $\delta\mathbf{v}(t,\mathbf{x})$ and $\delta T(t,\mathbf{x})$. In this work we are not interested in discussing the equations of fluctuating hydrodynamics including a fluctuating Fourier law, therefore we consider the hydrodynamics of a liquid at low temperatures, above the solidification temperature. 

As we discussed, to obtain the excitation spectra in liquids
on a macroscopic scale and over large time intervals one can start by discussing the hydrodynamic treatment of liquids, which is based on a continuum approximation with local conservation laws. Nonetheless, 
on a microscopic scale,
the nature of the vibration modes is determined by the  
interactions between its constituents, which need a quantum-mechanical description. 
For this system with a very large number of degrees of freedom, to study high-energy processes on the 
atomic/molecular-scale we use a formalism that unifies quantum mechanics with the classical theory of fields, i.e., quantum field theory.

With respect to these considerations, a remark is appropriate.
From the quantum Nyquist theorem, the spectral electromagnetic or scalar field density has a classical limit, where thermal fluctuations dominate, and a quantum regime at low temperatures, where quantum effects dominate \cite{weldon}. We get the spectral scalar field density:
\begin{equation}
    (\varphi^2)_\omega \propto \frac{\omega}{2} + \frac{\omega}{\exp{\left(\omega\beta\right)}-1},
\end{equation}
\noindent where $\beta =1/T$, in which $T$ is the temperature of the system. The quantum regime requires low temperatures and/or high frequencies, which is exactly the situation discussed in this work. The quantum fluctuations are dominant, and  
the liquid can be viewed as two weakly-coupled subsystems: phonons and the remainder of the liquid, as discussed by Andreev \cite{afandreev}. Because of these conditions, we assert that the noise field models quantum processes in the high-energy regime.

Instead of basing our discussion on a classical diffusion equation, we are interested in studying the emergent elementary excitations based on the quantization of the hydrodynamic fields. For a compressible fluid in thermodynamic equilibrium, the acoustic wave equation is obtained by linearizing the fluid dynamics equations for small disturbances around the constant equilibrium density and pressure. Thus, we have
\begin{align}
p(t,\mathbf{x})&=p_{0}+\delta p(t,\mathbf{x}),\\
\rho(t,\mathbf{x})&=\rho_{0}+\delta \rho(t,\mathbf{x}),\\
\bold{v}(t,\mathbf{x})&=\delta\bold{v}(t,\mathbf{x}),
\end{align}
where $\rho_{0}$ and $p_{0}$ are the constant equilibrium density and pressure, respectively. Assuming that the acoustic perturbation involves no rotational flow, we can write $\delta\bold{v}=\nabla \delta \chi$. Using the Euler and the mass balance equations and assuming that the acoustic perturbation is adiabatic, we obtain a linear, lossless wave equation for $\delta\rho(t,\mathbf{x})$ given by
\begin{equation}
\left(\frac{1}{u^{2}_{0}}\frac{\partial^{2}}{\partial t^{2}}-\Delta\right)\delta\rho(t,\mathbf{x})=0,
\end{equation}
and a similar equation for $\delta \chi(t,\mathbf{x})$. The constant $u_{0}$ is the longitudinal speed of sound. We would like to stress that a real liquid has finite viscosity and the liquid is not curl-free everywhere. In general acoustic processes, rotational effects are confined to the vicinity of the boundaries. 
Assuming an impenetrable boundary, we have the Neumann boundary conditions. We write $\bold{n}.\nabla\delta\rho(t,\mathbf{x})|_{\partial{V}}=0$. To study a simplified model, it is
convenient to leave aside viscosity effects and consider periodic boundary conditions. In this way, the translational invariance in the system is maintained.

The classical fields of the collective modes can be quantized. To proceed, let us discuss the quantization of the hydrodynamic fields. To define the elementary excitations of the acoustic waves, the sound quanta, we impose that the classical hydrodynamic fields $\delta \chi(t,\mathbf{x})$ and $\delta \rho(t,\mathbf{x})$ are Heisenberg operators obeying the equal-time commutation relations 
\begin{equation}
[\delta \chi(t,\mathbf{x}),\delta \chi(t,\mathbf{x}')]=[\delta \rho(t,\mathbf{x}),\delta \rho(t,\mathbf{x}')]=0
\end{equation}
and also  
\begin{equation}
[\delta \chi(t,\mathbf{x}),\delta \rho(t,\mathbf{x}')]=-i\delta(\mathbf{x}-\mathbf{x}').
\end{equation}
Using the noncommutativity algebra of the field operators and that the positive frequency modes associated with the hydrodynamic fields are given by $u_{\mathbf{p}}(t,\mathbf{x})$ where
\begin{equation}
u_{\mathbf{p}}(t,\mathbf{x}) = e^{i\big(\mathbf{p}\cdot\mathbf{x} -\omega(\bold{p}) t\big)},
\label{6}
\end{equation}
one can write the Fourier representation for the hydrodynamic field operators $\delta\rho(t,\mathbf{x})$ and $\delta \chi(t,\mathbf{x})$. They are given by 
\begin{equation}
\delta\rho(t,\mathbf{x})=\sum_{\bold{p}}i\left(\frac{\omega(\bold{p})}{2V}\right)^{\frac{1}{2}}\left(a_{\bold{p}}u_{\bold{p}}(t,\mathbf{x})-a^{\dagger}_{\bold{p}}u^{*}_{\bold{p}}(t,\mathbf{x})\right)
\label{7}
\end{equation} 
and
\begin{equation}
\delta\chi(t,\mathbf{x})=\sum_{\bold{p}}\left(\frac{1}{2\omega(\bold{p}) V}\right)^{\frac{1}{2}}\left(a_{\bold{p}}u_{\bold{p}}(t,\mathbf{x})+a^{\dagger}_{\bold{p}}u^{*}_{\bold{p}}(t,\mathbf{x})\right),
\label{8}
\end{equation}
where $a_{\bold{p}}$ and $a^{\dagger}_{\bold{p}}$ are annihilation and creation operators of elementary excitations with angular frequency $\omega(\bold{p})$ and 
pseudo-momentum $\bold{p}$.
We assume that the phonon angular frequency is written as $\omega(\bold{p})=u_{0}|\bold{p}|$ \cite{abri,lan}. This linear dispersion relation is a reasonable approximation for phonon wavelengths much longer than the liquid intermolecular distance, and satisfies the condition 
\begin{equation}
\lim_{\bold{p}\rightarrow 0}\omega(\bold{p})=0.
\end{equation}
To proceed, we have to implement the physical condition of the Wightman axioms: the states of this physical system are realized as elements of a Hilbert space \cite{wightman}. 
The construction of the Hilbert space of multi-quasi-particle states is straightforward. The state without elementary excitations is the Fock vacuum state $|\Omega_{0}\rangle$ of the phononic field. It is defined using that $a_{\bold{p}}|\,\Omega_{0}\rangle=0 ~\forall ~\bold{p}$. 
All the excited states can be created by acting on the Fock vacuum state with the local operators $a^{\dagger}_{\bold{p}}$ and $a_{\bold{p}}$. An arbitrary state of the Hilbert space is given by a linear superposition of  multi-elementary excitation states. It can be represented as
\begin{equation}
|\Psi\rangle =\sum_{q=0}^{\infty}\frac{1}{(q!)^{\frac{1}{2}}}\int\psi_{q}({\bold{p}}_{1},\dots,{\bold{p}}_{q})\,a^{\dagger}_{\bold{p}_{1}}...
a^{\dagger}_{\bold{p}_{q}}\prod_{i=1}^{q}(d^{3}p_{i})|~\Omega_{0}\rangle,
\end{equation}
where $\psi_{0} \in \mathbb{C}$ and $\psi_{q}({\bold{p}}_{1},...{\bold{p}}_{q})$ for $q\geq 2$ are symmetric functions. We have
\begin{equation}
\langle \Psi|\Psi\rangle=\sum_{q=0}^{\infty}\int|\psi_{q}({\bold{p}}_{1},...{\bold{p}}_{q})|^{2}\prod_{i=1}^{q}(d^{3}p_{i})<\infty.
\end{equation} 
The above representation defines the Fock space of the system. We define also the causal two-point correlation function for the phononic field as
\begin{equation}
G^{(2)}(t,\mathbf{x};t',\bold{x}')=-i\langle \Omega_{0}~|T\left[\delta\rho(t,\mathbf{x})~ \delta\rho(t',\mathbf{x}')\right]|~\Omega_{0}\rangle, 
\label{nova11}
\end{equation}
where $T[...]$ is the Dyson time-ordered product.
Substituting the Fourier representation of the field operator $\delta\rho(t,\mathbf{x})$, defined in Eq.\eqref{7}, in Eq.\eqref{nova11}, one obtains that the causal correlation function can be written as
\begin{eqnarray}
&&G^{(2)}(t,\mathbf{x};t',\mathbf{x}')=\nonumber\\
&&-\frac{i}{V}\sum_{\bold{p}}\frac{\omega(\bold{p})}{2}\left(\theta(t)~u_{\bold{p}}(t,\mathbf{x})+\theta(-t)~u^{*}_{\bold{p}}(t,\mathbf{x})\right),
\label{42}
\end{eqnarray}
where $\theta(x)$ is the Heaviside step function. 
The Fourier representation of the causal correlation function of the phonons can be readily derived.
It can be written as  
\begin{equation}
\bar{G}^{(2)}(\upsilon,\bold{p})=\frac{\omega^{2}(\bold{p})}{\upsilon^{2}-\omega^{2}(\bold{p})+i\delta},
\label{431}
\end{equation}
where the infinitesimal term in the denominator indicates in which half-plane of the complex frequency the corresponding integrals will converge.
In the following, we use both formalisms, the canonical and the functional formalism, concomitantly. 

The action functional for a phononic field in a liquid or in a solid at moderate temperature with some ordered structure is given by 
\begin{equation}
S(\delta\rho)=S_{0}(\delta\rho)+S_{int}(\delta\rho)
\end{equation}
where the second nonlinear contribution must be a polynomial of the field. It appears if we quantize the acoustic waves in solids, where the quantization of the classical fields of the collective modes includes one longitudinal and two transverse acoustic modes, and the presence of anharmonicity introduces phonon-phonon interactions, resulting in the Landau-Rumer finite lifetime of such excitations \cite{rumer,peter,in}. The free action functional of the liquid is written as
\begin{equation}
S_{0}(\delta\rho)=\frac{1}{2}\int d^{4}x
~\left[\delta\rho(t,\mathbf{x})\left(\frac{1}{u^{2}_{0}}\frac{\partial^{2}}{\partial t^{2}}-\Delta\right)\delta\rho(t,\mathbf{x})\right],
\label{9}
\end{equation}
where $\Delta$ is the Laplace operator which acts on scalar functions defined in a finite time interval $([t_{a},t_{b}])$ and in $V \subset \mathbb{R}^{3}$, and $u_{0}$ is the longitudinal speed of the sound wave. We assume that $t_{b}-t_{a}\gg \frac{V^{\frac{1}{3}}}{u_{0}}$. The eigenfunctions of $(-\Delta)$ form a complete basis in the function space $L_{2}(V)$ of measurable and square-integrable functions on $V$. In the following, we discuss the functional approach that can be used to describe the propagation of quantized acoustic waves, i.e., sound quanta in the liquid.

The functional integral representation for the vacuum persistence functional of the scalar field theory in the presence of an external scalar source $j(t,\mathbf{x})$ is given by the functional integral 
\small\begin{equation}
Z(j)=\mathcal{N}\int \mathcal{D}\delta\rho ~ \exp\left(iS(\delta\rho)+i\int d^{4}x ~ j(t,\mathbf{x})~\delta\rho(t,\mathbf{x})\right),
\label{11}
\end{equation}
\normalsize where $\mathcal{D}\delta\rho$ denotes integration over all functions $\delta\rho(t,\mathbf{x})$ of space and time, and $\mathcal{N}$ is the normalization factor, using that $Z(j)|_{j=0}=1$. Since $S(\delta\rho)$ is the action integral for the classical field theory, the functional integrals are over all classical field histories.
The functional $Z(j)$ is the generating functional of the vacuum expectation values of chronologically-ordered products of the field operators.
 Note that the $Z(j)|_{j=0}$ has a purely formal meaning since, even using the normal ordering $:S_{int}(\delta\rho):$, we have a kind of a complex measure in the function space, i.e., $
\mathcal{D}\mu^2(\delta\rho)=\mathcal{N}\exp{(iS(\delta\rho))}\mathcal{D}\delta\rho$.

The usefulness of $Z(j)$ is that it permits one to construct the correlation functions, i.e., the vacuum expectation values of chronologically-ordered products of the  
field operators, by performing a suitable number of functional differentiations with respect to the source. For an arbitrary theory with interaction action $S_{int}(\delta\rho)$ and $G_{0}(t,\mathbf{x};t',\mathbf{x}')$, the free two-point correlation function, we construct a perturbative theory (the Feynman-Dyson series) by writing the generating functional as 
\small\begin{align}
&Z(j)=\mathcal{N}\exp\left[i\int d^{4}x ~S_{int}\left(\frac{1}{i}\frac{\delta}{\delta j(t,\mathbf{x})}\right)\right]  \nonumber \\ 
&\times \exp\left[\frac{i}{2}\int d^{4}x\int d^{4}x'~j(t,\mathbf{x})~G_{0}(t,\mathbf{x};t',\mathbf{x}')~j(t',\mathbf{x}') \right]. \label{12}
\end{align}

\normalsize The coefficients of the expansion of $Z(j)$ in a Taylor functional series in $j(t,\mathbf{x})$ determine the correlation functions of the model. The perturbative theory is obtained by expanding $Z(j)$ in powers of the coupling constant. The correlation functions are given by the sum of all diagrams with $n$ external legs, including the disconnected diagrams. The vacuum diagrams are cancelled by the normalization factor. 

\section{The phononic field theory in a fluctuation environment}\label{sec:disoderedLG}

Let us start discussing the Frenkel approach for short time processes in liquids accessing the solid-like regime. Frenkel defines the liquid relaxation time $\tau_{f}$, and a critical angular frequency defined as $\omega_{F}=\frac{2\pi}{\tau_{f}}$. For times shorter than  the liquid relaxation time $\tau_{f}$, the local structure of the liquid remains static, similar to that of a solid. Therefore, for times shorter than $\tau_{f}$, i.e., high frequencies $\omega \geq \omega_{F}$,the system supports one longitudinal mode and two transverse modes. The dispersion relation obtained is 
\begin{equation}
\omega(\bold{p})=-\frac{i}{2\tau_{f}}+
\left(u^{2}\bold{p}^{2}-\frac{1}{4\tau_{f}^{2}}\right)^\frac{1}{2},
\label{frenkel}
\end{equation}
where $u$ is the transverse speed of sound. There is a critical value of the pseudo-momentum where we have propagating modes.
This dispersion relation characterizes a solid-like elastic regime in liquids. 

Here, we develop an alternative approach to obtain results similar to those found in the literature. Our starting point is an effective model of the non-hydrodynamic degrees of freedom. In the kinetic regime, instead of using Frenkel's ideas or discussing the microscopic equations of motion at the molecular scale, we are introducing an additive quantum noise field in the model, a randomly fluctuating environment. We discuss free theories, integrating out the noise in the augmented generating functional of connected correlation functions. The problem raised is how the quantized hydrodynamic fields change after integrating out the noise field. Since noise also induces local fluctuations in the quantum fields, after this procedure, we are analyzing noise-induced effects on  quantized acoustic perturbations. We consider a noise field $h(t,\mathbf{x})$ as a function of time and points in a finite volume $V \subset \mathbb{R}^{3}$. Since the perturbation theory is based on the free two-point correlation function, we discuss only the free field theory.

Let us discuss the introduction of the solid-like regime. Consider a low-temperature liquid, where the quantum fluctuations are dominant over the temperature fluctuation. In this case, there is no state with thermally excited particles. We start from the Navier-Stokes equation, with the coefficients of bulk and shear viscosity. Using a  linearized equation, an adiabatic assumption,  and also a linearized equation of continuity, one obtains a lossy wave equation \cite{lamb}. For the linearized case the differential equations for the sound waves in viscous media are given by
\begin{equation}
\frac{\partial^{2}\boldsymbol{\vartheta}_{\lambda}(t,\mathbf{x})}{\partial t^{2}}=\Delta\left(c_{\lambda}^{2}+D_{\lambda}\frac{\partial}{\partial t}\right){\boldsymbol{\vartheta}}_{\lambda}(t,\mathbf{x}).
\label{navier}
\end{equation}
The sound waves $\boldsymbol{\vartheta}_\lambda(t,x)$ have two components: longitudinal and transverse, defined as $\boldsymbol{\vartheta}_l(t,\mathbf{x})$ and $\boldsymbol{\vartheta}_t(t,\mathbf{x})$, respectively. In the above equation, $c_{\lambda}$ and $D_{\lambda}$ are the speed of propagation and a parameter proportional to the diffusion constant of the $\lambda$ branch, respectively. The subscript $\lambda$ also refers to the longitudinal and transverse displacement fields.  

 Here we assume the following hypotheses: (i) for a short time scale, any liquid behaves like a solid, and is able to oppose tangential stresses, with the presence of transverse acoustic modes, and (ii)  to take into account the molecular environment, we use a random noise $\bold{h}_\lambda(t,\mathbf{x})$. 
Using (i) and (ii) the acoustic perturbations are described by the following wave equations:
\begin{equation}
	\left(\frac{1}{c_\lambda^2}\frac{\partial^2}{\partial t^2} - \Delta \right)\boldsymbol{\vartheta}_\lambda(t,\mathbf{x}) + \bold{h}_\lambda(t,\mathbf{x}) = 0,
\end{equation}
where again $c_\lambda$ are the sound speeds and $\boldsymbol{\vartheta}_\lambda$ are the elastic waves, i.e., displacements of the ``solid structure", as discussed in Ref. \cite{ela}. Here we use $c_l=u_{0}$ and $c_t=u$ as the longitudinal and transverse sound speeds, respectively. These quantities will be used in the action functional that describes the system in this solid-like regime as the sum of both components, i.e., $S_0 = S_l + S_t$, where 
\begin{align}
S_\lambda(\boldsymbol{\vartheta}_\lambda, \bold{h}_\lambda) = \int d^4 x &\left[ \frac{1}{2}\boldsymbol{\vartheta}_\lambda(t,\bold{x})\cdot\left(\frac{1}{c_\lambda^2}\frac{\partial^2}{\partial t^2} - \Delta \right) \boldsymbol{\vartheta}_\lambda(t,\bold{x}) \right. \nonumber\\
& +  \bold{h}_\lambda(t,\mathbf{x}) \cdot\boldsymbol{\vartheta}_\lambda(t,\bold{x})\bigg],\label{eq: l and t modes}
\end{align}
in which $\lambda = l,t$. As usual, we define the $e^{(\lambda)}_{i}$ as the transverse and longitudinal polarization vectors, where $p^{i}e_{i}^{(\lambda)}=0$, for $\lambda=2,3$ and $p^{i}e_{i}^{(\lambda)}=|\bold{p}|$, for $\lambda=1$.  We have $e^{(1)}_{i}=(1,0,0)$, $e^{(2)}_{i}=(0,1,0)$ and $e^{(3)}_{i}=(0,0,1)$. Note that we are assuming that different polarizations are decoupled from each other. 

In the liquid, this assumption can be used. Also, since at low temperatures there is an attenuation of the longitudinal, high-frequency phonons \cite{orbach}, here we discuss only the transverse displacements. Therefore, we can write the action functional for each transverse component, where we are writing, for simplicity, $\varphi(t,\mathbf{x})=\vartheta_t(t,\bold{x})$ and $h(t,\mathbf{x})=h_{t}(t,\mathbf{x})$ for the transverse components. The free action functional for each transverse degree of freedom with the contribution of the noise field can be defined by  
\small\begin{eqnarray}
&&S(\varphi, h)=\nonumber\\
&&\int\,d^{4}x~\left[\frac{1}{2}\varphi(t,\mathbf{x})\left(\frac{1}{u^{2}}\frac{\partial^{2}}{\partial t^{2}}-\Delta\right)\varphi(t,\mathbf{x}) + h(t,\mathbf{x})~\varphi(t,\mathbf{x})\right].\nonumber\\
\end{eqnarray}

\normalsize The basic idea is that we are considering an augmented functional\footnote{Instead of using the terminology noisy functional we prefer to use augmented functional, i.e., a functional of the noise field and also of the source.} $Z(j,h)$ i.e., the usual generating functional of $n$-point correlation functions of the model augmented by an additive white-noise field. Note that in the path integral formalism the formal oscillatory behavior of the integrand leads us to conclude that the sum over field configurations is dominated by the field configuration of stationary phase, i.e., the solution to the classical field equation.
The functional integral representation of this augmented functional is 
\small\begin{equation}
Z(j,h)=\mathcal{N}'\int \mathcal{D}\varphi~ \exp\left(i\,S(\varphi,h)+i\int d^{4}x~j(t,\mathbf{x})~\varphi(t,\mathbf{x})\right).\label{15}
\end{equation}
\normalsize The $Z(j,h)$ corresponds to the functional integral $Z(j)$ in the presence of the noise field.
There are some similarities between 
our approach and the one used in a purely classical scenario studying the 
functional formulation of the problem of turbulence, where the classical fluid described by a Navier-Stokes equation is under the effect of a random force \cite{hopf}. Since we have that $Z(j,h)|_{h=0}=Z(j)$, 
and also that $Z(j,h)|_{j=0}=Z(h)$, this augmented functional also satisfies that $Z(h)|_{h=0}=1$.

To define a generating functional of correlation functions in the field theory with the presence of the noise, we define 
\begin{equation}
\mathcal{D}\chi=\mathcal{N}''e^{iC(h)}\mathcal{D}h,
\label{dx}
\end{equation}
where
\begin{equation}
C(h)=-\frac{1}{2\,\sigma^{2}}\int d^{4}x~\left( h(t,\mathbf{x})\right)^{2}.
\label{16}
\end{equation}
The $\mathcal{N}''$ is an immaterial normalization factor that will be omitted in subsequent calculations, and $\mathcal{D}h$ is a purely formal notation.  

To proceed, one can consider another augmented generating functional, i.e.,  
the usual generating functional of connected correlation functions defined with the noise field, $i.e.,$ the augmented functional $W(j,h)=-i\log Z(j,h)$, also for a specific configuration in functional space of the noise field. First, integrating out the noise, using Eqs. \eqref{dx} and \eqref{16} we define the new functional $\mathbb{Q}\left[W(j,h)\right]$ as 
\begin{equation}
\mathbb{Q}\left[W(j,h)\right]= \int \mathcal{D}\chi ~W(j,h).
\label{19}
\end{equation}
\iffalse
Taking the average of a random variable over the ensemble of realizations or integrating out the noise field using the functional integral formalism are conceptually different procedures. We will now turn to interpretation issues. When we average over a random variable we use the symbol $\mathbb{E}\bigl[...\bigr]$. The average of a random variable is used in statistical field theory and also in Euclidean field theory, with analytically continued vacuum expectation values of product of field operators. The Euclidean program suggests that many problems in field theory are really problems in probability theory. In systems governed by classical physics, the formalism is based on a measure space, i.e., a set $X$ together with a sigma-algebra of subsets of $X$ and a measure defined in that algebra, i.e., a non-negative and countable additive set function. A real random variable is a measurable real value function on $X$. Since the noise field $h(t,\mathbf{x})$ is not a random variable in the formal sense, integrating out the noise in some functional we use the symbol $\mathbb{Q}\bigl[...\bigr]$.
\fi
We can consider $\mathbb{Q}\bigl[...\bigr]$ as the ``expectation value" of a functional over some specific complex measure. In the path integral formalism $Z(j)$ is the vacuum persistence functional in the presence of a scalar source, i.e., a functional integral over all classical field histories. Discussing an effective model on the molecular-scale description, the Eq. \eqref{19} is not a functional integral over the noise field histories. It is only a functional integral over all configuration space of the noise field, ``averaging" $W(j,h)$. We will show that it is possible to write $\mathbb{Q}\bigl[...\bigr]$ as a functional series, with effective actions. For each term of the series there is a functional integral over all classical fields histories, evaluated for finite temporal intervals.  

Another problem arises in close connection with the above remarks. Working with $\mathbb{Q}\left[Z(j,h)\right]$ it is possible to show that the effects of the noise field are to generate only one tachyonic-like field in the liquid. The energy spectrum of the elementary quasi-particles must contain phononic and quasi-particles with gaps in the pseudo-momentum space. Strictly speaking, in the kinetic regime, a rather natural way to find modified dispersion relations for the elementary excitations of the liquid is to perform the functional integral of the augmented generating functional $W(j,h)$ i.e.,  evaluating $\mathbb{Q}\left[W(j,h)\right]$. This procedure to work with $\mathbb{Q}\left[W(j,h)\right]$ is similar to the one used in statistical field theory, but with a substantially different interpretation. As we discussed, the Eq. \eqref{19}  means a functional integral over all noise field configurations in a function space. In the theory of classical random fields, with randomness and competing interactions, the free energy must be self-averaging over all the realizations of the random variable i.e., performing $\mathbb{E}\left[W(j,h)\right]$. 

There is another aspect in our problem which has to be considered. There are different approaches in the literature, discussing systems with quenched disorder in statistical mechanics and also statistical field theory, to integrate out the disorder to obtain $\mathbb{E}\bigl[W(j,h)\bigr]$. One of them is the replica ``trick'' \cite{livro1,livro3}. For other approaches see Refs.\cite{dominicis,zip,livro4,efe1}. We have to integrate out the noise in the augmented generating functional $W(j,h)$. For a generic $C(h)$ we define the functional integral
\begin{equation}
\mathbb{Q}\left[W(j,h)\right]=-i\int\mathcal{D}\chi ~\log Z(j,h).
\label{24}
\end{equation}
If we are able to find a theoretically tractable expression for $\mathbb{Q}\left[W(j,h)\right]$, we are able to determine how the presence of the noise affects the behavior of the noiseless system, obtaining the excitation spectra of the liquid.

\section{Phononic quasiparticles and excitations with gaps in the pseudo-momentum space}
\label{sec:thermal mass}

An alternative method 
that has been discussed in the literature to represent $\mathbb{E}\left[W(j,h)\right]$ in a tractable way is the distributional zeta-function method \cite{distributional,distributional2,zarro1,zarro2,polymer,soares,spin-glass,non}. 

Here we adapt the method for the case of path integrals including the noise field. Using the augmented functional integral $Z(j,h)$ given by Eq.\eqref{15}, the distributional zeta-function, $\Phi(s)$, becomes a configurational zeta-function, and is defined formally as
\begin{equation}
\Phi(s)=\int \mathcal{D}\chi~Z(j,h)^{-s},
\label{25}
\end{equation}
for $s\in \mathbb{C}$. A new generating functional given by Eq.\eqref{24} tracing out the additive noise can be written as 
\begin{equation}
\mathbb{Q}\left[W(j,h)\right]=i\frac{d}{ds}\Phi(s)\Bigg|_{s=0^{+}}, ~\text{Re}(s) \geq 0,  
\end{equation}
where one defines the complex exponential $n^{-s}=\exp(-s\log n)$, with $\log n\in\mathbb{R}$. 
Using analytic tools, we define a new generating functional $\mathbb{Q}\left[W(j,h)\right]$, where the noise field was integrated out. 
The integrated generating functional can be represented as
\small\begin{eqnarray}
&&\mathbb{Q}\left[W(j,h)\right]=\nonumber\\
&&i\sum_{k=1}^{\infty} \frac{(-1)^{k}a^{k}}{k k!}~
\mathbb{Q}\left[Z(j,h)^{k}\right]+i\log(a)-i\gamma-iR(a,j),\nonumber\\
\label{m23e}
\end{eqnarray}
\normalsize where $a$ is a dimensionless arbitrary constant, $\gamma$ is the Euler-Mascheroni constant, $R(a)$, given by
\begin{equation}
R(a,j)=-\int \mathcal{D}\chi\int_{a}^{\infty}\frac{dy}{y}~\exp\left[-Z(j,h)y\right].
\end{equation}
In the functional series we have
\begin{equation}
\mathbb{Q}\left[Z(j,h)^{k}\right]=\int \mathcal{D}\chi~
Z(j,h)^{k}.
\label{19a}
\end{equation}
For large $a$, $|R(a)|$ is quite small, therefore, the dominant contribution to the integrated generating functional is given by the ``moments" of the generating functional of correlation functions of the model. We get 
\begin{equation}
\mathbb{Q}\left[W(j,h)\right]=i\sum_{k=1}^{\infty} \frac{(-1)^{k}}{k k!}~
\mathbb{Q}\left[Z(j,h)^{k}\right],
\label{29}
\end{equation}
where the $a$ constant was absorbed in the functional measures. Using that $\Box=\Bigl(\frac{1}{u^{2}}\frac{\partial^{2}}{\partial t^{2}}-\Delta\Bigr)$
we define the following constants 
$\mathcal{N}_{0}(k)$, and $\mathcal{N}_{l}^{(k)}$ given by 
\begin{eqnarray}
&&\frac{1}{\mathcal{N}_{0}(k)}=\nonumber\\ &&\int\mathcal{D}\psi^{(k)}
~\exp\left[\frac{i}{2}\int d^{4}x~\psi^{(k)}(t,\mathbf{x})\left(\Box-k\sigma^{2}\right)\psi^{(k)}(t,\mathbf{x})\right]\nonumber\\
\end{eqnarray}
and for $l\geq 2$
\begin{equation}
\frac{1}{\mathcal{N}_{l}^{(k)}}
= \int\mathcal{D}\phi^{(k)}_l ~
\exp\left[\frac{i}{2}\int d^{4}x~\phi^{(k)}_{l}(t,\mathbf{x})~\Box~\phi^{(k)}_{l}(t,\mathbf{x})\right].
\label{34b}
\end{equation}
Each term of the functional series is written as a product of $k$ functional integrals. We have
\begin{eqnarray}
&&\mathbb{Q}
\left[Z(h,j)^{k}\right]\Bigg|_{k=1} = \nonumber\\
&&\mathcal{N}_0(k)\int\mathcal{D}\psi^{(k)}~\exp\left[i S_2\left(\psi^{(k)},j^{(k)}_\psi\right)\right]\Bigg|_{k=1}\label{30a}
\end{eqnarray}
and
\begin{eqnarray}
&&\mathbb{Q}
\left[Z(h,j)^{k}\right]\Bigg|_{k\geq 2}=\nonumber\\
&&\mathcal{N}_0(k)\int\mathcal{D}\psi^{(k)}~\exp\left[iS_2\left(\psi^{(k)},j^{(k)}_\psi\right)\right] \nonumber \\
&&\times \mathcal{N}^{(k)}_\phi\int\prod_{l=2}^{k}\mathcal{D}\phi^{(k)}_l~\exp\left[iS^{(k)}_1\left(\phi^{(k)}_{l},j^{(k)}_{\phi_l}\right)\right],
\label{30}
\end{eqnarray}
where $\mathcal{N}^{(k)}_\phi=\prod_{l=2}^{k}\mathcal{N}_l^{(k)}$ and $\mathcal{D}\phi^{(k)}_l$ are products of integration over all functions $\phi^{(k)}_{l}(t,\mathbf{x})$ of space and time. The notation $\phi^{(k)}_{l}$ means that we are considering in the $k$-term of the functional series, the $l$-th component of the multiplet with $k$ components. In this case, the new effective actions $S^{(k)}_{1}\left(\phi^{(k)}_{l},j^{(k)}_{\phi_l}\right)$ and $S_2\left(\psi^{(k)},j^{(k)}_\psi\right)$
are written as
\small\begin{align}
&S^{(k)}_{1}\left(\phi^{(k)}_{l},j^{(k)}_{\phi_l}\right)\nonumber\\
&=\sum_{l=2}^{k}\int\ d^{4}x ~\left[\phi^{(k)}_{l}(t,\mathbf{x})~j^{(k)}_{\phi_l}(t,\mathbf{x})+\tfrac{1}{2}\phi^{(k)}_{l}(t,\mathbf{x})~\Box~
\phi^{(k)}_{l}(t,\mathbf{x})\right]
\label{31}
\end{align}
\normalsize and
\small\begin{align}
& S_2\left(\psi^{(k)},j^{(k)}_{\psi}\right)\nonumber \\
&=\int d^{4}x~\left[\psi^{(k)}(t,\mathbf{x})j^{(k)}_\psi(t,\mathbf{x})+\frac{1}{2}\psi^{(k)}(t,\mathbf{x})~\Box~\psi^{(k)}(t,\mathbf{x})\right]. 
\label{32}
\end{align}
\normalsize From the above equation it is clear that integrating the noise field in the generating functional $Z(j,h)$, defining  $\mathbb{Q}\left[Z(j,h)\right]$ is exactly the term $k=1$ in the functional series. In conclusion, the effects of the noise field in the liquid are to generate a field with a dispersion relation with a gap in the pseudo-momentum space.

Now, let us define the functionals ${Z}_\psi^{(k)}\left(j^{(k)}_\psi\right)$ and ${Z}_{\phi_{l}}^{(k)}\left(j^{(k)}_{\phi_l}\right)$ as 
\begin{equation}
{Z}_\psi^{(k)}\left(j^{(k)}_\psi\right) = \mathcal{N}_0(k)\int\,\mathcal{D}\psi^{(k)}\,\exp\Bigl[i S_2\left(\psi^{(k)},j^{(k)}_\psi\right)\Bigr]
\end{equation}
and for $l\geq 2$
\begin{equation}
{Z}_{\phi_l}^{(k)}\left(j^{(k)}_{\phi_l}\right) = \mathcal{N}_l^{(k)} \int\mathcal{D}\phi^{(k)}_l~ \exp{\left[iS^{(k)}_3\left(\phi^{(k)}_l, j^{(k)}_{\phi_l}\right)\right]}
\end{equation}
where $S^{(k)}_3\left(\phi^{(k)}_l, j^{(k)}_{\phi_l}\right)$ is the action for the $l$-th phononic field in the $k-$th term of the series. We have obtained that each term of the functional series defined by Eq.\eqref{29} is  represented by  products of $k$ functional integrals, one of them for the k-gap field and the others for the $k-1$ functional integrals for the phononic fields. All the mathematical tools needed to obtain these results were developed in earlier papers \cite{PRE,PRL,PRD,analogwormhole}. We adopt a modified functional integral in order to proceed using Eqs. \eqref{30a}-\eqref{32}. In the $k$-th term of the series, we are using the notation $\psi_{k}(t,\mathbf{x})$, in order to specify this particular field. We impose that this field $\psi_{k}(t,\mathbf{x})$ is assumed  to possess non-vanishing Fourier components only for $\bold{p}^{2}\geq k\sigma^{2}$. This is exactly the approach used by  Feinberg, Arons and Sudarshan in the canonical quantization of tachyons to avoid imaginary frequencies \cite{t2,t3}. Since both effective actions defined by Eq.\eqref{31} and Eq.\eqref{32} describe free quantum field theories, all the $2n$-point correlation functions of the model are written in terms of the two-points correlation functions associated with the phononic and the k-gap fields.  

\section{The two-point correlation functions for phononic and k-gap fields}\label{sec:two}

The aim of this section is to obtain the two-point correlation functions for the phononic and the k-gap fields. Since the scalar source $j(t,\mathbf{x})$ was introduced as a device to define the generating functionals of the theory, one has the freedom to choose a particular source distribution. This point will be clarified later. It is also convenient to define 
the quantity
\begin{equation}
\mathbb{Q}\left[Z(h,j)^{k}\right]\Big|_{k=1}=
{Z}_\psi^{(1)}\left(j^{(1)}_\psi\right)
\end{equation}
and
\begin{equation}
\mathbb{Q}\left[Z(h,j)^{k}\right]\Bigg|_{k \geq 2}=
{Z}_\psi^{(k)}\left(j^{(k)}_\psi\right) \prod_{r=2}^{k}{Z}^{(k)}_{\phi_r}\left(j^{(k)}_{\phi_r}\right).
\label{sim2}
\end{equation}
To proceed, let us define the coefficient $c^{(1)}_{k}=\frac{(-1)^{k}}{kk!}$. Substituting Eq.\eqref{sim2} in Eq.\eqref{29} the functional series is written as 
\begin{eqnarray}
&&\mathbb{Q}\,\bigl[(W(h,j))\bigr] = \nonumber\\
&&ic_1^{(1)}{Z}_\psi^{(1)}(j^{(1)}_\psi)+i\sum_{k=2}^{\infty}c_{k}^{(1)}\,
{Z}_\psi^{(k)}\left(j^{(k)}_\psi\right) \prod_{r=2}^{k}{Z}^{(k)}_{\phi_r}\left(j^{(k)}_{\phi_r}\right).\nonumber\\
\label{29a}
\end{eqnarray}
As usual, to make contact with the two-point correlation functions of the model we must perform two functional derivatives with respect to the sources of the model. We have
\begin{align}
&\frac{\delta^{2}
\mathbb{Q}\left[(W(h,j))\right]}{\delta j^{(k)}_{\psi}(t,\mathbf{x})\delta j^{(k)}_{\psi}(t',\mathbf{x}')}\Bigg|_{j^{(k)}_{\psi}=0}=\nonumber \\
&ic_{1}^{(1)}~
\frac{\delta^{2}{Z}_{\psi}^{(1)}\left(j^{(k)}_{\psi}\right)}{\delta j^{(k)}_{\psi}(t,\mathbf{x})\delta j^{(k)}_{\psi}(t',\mathbf{x}')}\bigg|_{j^{(k)}_{\psi}=0}+\nonumber\\
&i\prod_{r=2}^{k}{Z}^{(k)}_r\left(j^{(k)}_{\phi_r}\right)\sum_{k=2}^{\infty}c_{k}^{(1)}~\frac{\delta^{2}{Z}_{\psi}^{(k)}\left(j^{(k)}_{\psi}\right)}{\delta j^{(k)}_{\psi}(t,\mathbf{x})\delta j^{(k)}_{\psi}(t',\mathbf{x}')}\bigg|_{j^{(k)}_{\psi}=0},
\label{29ab}
\end{align}
and
\begin{align}
&\frac{\delta^{2}
\mathbb{Q}\,\bigl[(W(h,j))\bigr]}{\delta j^{(k)}_{\phi_l}(t,\mathbf{x})\delta j^{(k)}_{\phi_l}(t',\mathbf{x}')}\bigg|_{j^{(k)}_{\phi_{l}}=0}=\nonumber \\ 
&i\sum_{k=2}^{\infty}c_{k}^{(1)}\,{Z}_{\psi}^{(k)}(j_{\psi})\prod_{r=2}^{k}\frac{\delta^{2}{Z}_{\phi_r}^{(k)}(j^{(k)}_{\phi_r})}{\delta j^{(k)}_{\phi_{l}}(t,\mathbf{x})\delta j^{(k)}_{\phi_{l}}(t',\mathbf{x}')}\bigg|_{j^{(k)}_{\phi_{l}}=0}.
\label{29ab2}
\end{align}
Using the fact that ${Z}_{\phi_l}^{(k)}\left(j^{(k)}_{\phi_l}\right)|_{j^{(k)}_{\phi_l}=0}=1$ and also that ${Z}_\psi^{(k)}\left(j^{(k)}_{\psi}\right)|_{j^{(k)}_{\psi}=0}=1$ we define two distinct functionals and  functional series. We can write two functional series, using Eq.\eqref{29a}. They are 
\begin{equation}
\mathbb{Q}[W(h,j)]\Bigg|_{j^{(2)}_{\phi_l}=j^{(3)}_{\phi_l}=\dots=j^{(k)}_{\phi_l}
=0} = i\sum_{k=1}^{\infty}c_{k}^{(1)}
{Z}_\psi^{(k)}\left(j^{(k)}_{\psi}\right),
\label{29aa}
\end{equation}
for all $l$ we have $j^{(k)}_{\phi_l}=0$ and 
\begin{equation}
\mathbb{Q}[W(h,j)]\Bigg|_{j^{(1)}_{\psi}=j^{(2)}_{\psi}=\dots=j^{(k)}_{\psi}=0}
=i\sum_{k=2}^{\infty}c_{k}^{(1)}\prod_{r=2}^{k}
{Z}_{\phi_r}^{(k)}\left(j^{(k)}_{\phi_r}\right).
\label{29aa2}
\end{equation}
In a noiseless system one defines the correlation functions and the connected correlation functions. These correlation functions can be defined by performing a functional expansion of the generating functional of correlation and connected correlation functions respectively. It is to be noted that the same construction  can be done for a system under the effects of an additive noise. Let us define the following functional series. We have
\small\begin{align}
{Z}_\psi^{(k)}\left(j^{(k)}_{\psi}\right)
=\sum_{n=1}^{\infty}\frac{i^{n-1}}{n!}\int \prod_{s=1}^{n}d^{4}x_{s}~j^{(k)}_{\psi}(t_{1},\mathbf{x}_{1})\dots j^{(k)}_{\psi}(t_{n},\mathbf{x}_{n})\nonumber\\
\times {G}^{(n)}_{\psi^{(k)}}\left(t_{1},\mathbf{x}_{1};\dots;t_{n},\mathbf{x}_{n},k\right),
\label{18}
\end{align} 
\normalsize and for the $l$-th phononic component of the $k$-th multiplet we have
\small\begin{align}
{Z}_{\phi_{l}}^{(k)}\left(j^{(k)}_{\phi_l}\right)
= \sum_{n=1}^{\infty}\frac{i^{n-1}}{n!}\int \prod_{s=1}^{n}d^{4}x_{s}~j^{(k)}_{\phi_l}(t_{1},\mathbf{x}_{1})\dots j^{(k)}_{\phi_l}(t_{n},\mathbf{x}_{n})\nonumber\\
\times{G}^{(n)}_{\phi^{(k)}_l}\left(t_{1},\mathbf{x}_{1};\dots;t_{n},\mathbf{x}_{n}\right).
\label{18a}
\end{align} 

\normalsize From here, what follows is straightforward. Using functional derivatives we obtain the $n$-point correlation functions ${G}^{(n)}_{\phi^{(k)}_l}\bigl(t_{1},\mathbf{x}_{1};\dots;t_{n},\mathbf{x}_{n}\bigr)$ and  ${G}^{(n)}_{\psi^{(k)}}\bigl(t_{1},\mathbf{x}_{1};\dots;t_{n},\mathbf{x}_{n}\bigr)$ of the model, i.e. the original connected correlation functions modified by the effects of the noise field. We have
\begin{eqnarray}\label{23}
 &&   \frac{\delta^{n}
{Z}_\psi^{(k)}\left(j^{(k)}_{\psi}\right)}{\delta j^{(k)}_{\psi}(t_{1},\mathbf{x}_{1})\dots\delta j^{(k)}_{\psi}(t_{n},\mathbf{x}_{n})}\Bigg|_{j^{(k)}_{\psi}=0}=\nonumber\\
&&i^{n-1}{G}_{\psi^{(k)}}^{(n)}(t_{1},\mathbf{x}_{1};\dots;t_{n},\mathbf{x}_{n};k)
\end{eqnarray}
and
\begin{eqnarray}\label{23a}
&&\frac{\delta^{n}
{Z}_{\phi_{l}}^{(k)}\left(j^{(k)}_{\phi_l}\right)
}{\delta j^{(k)}_{\phi_l}(t_{1},\mathbf{x}_{1})\dots\delta j^{(k)}_{\phi_l}(t_{n},\mathbf{x}_{n})}\bigg|_{j^{(k)}_{\phi_l}=0}=\nonumber\\
&&i^{n-1}{G}_{\phi^{(k)}_l}^{(n)}(t_{1},\mathbf{x}_{1};\dots;t_{n},\mathbf{x}_{n}).
\end{eqnarray}
Let us define $G_{\psi^{(k)}}^{(2)}(t,\mathbf{x};t',\mathbf{x}')$ as the causal two-point correlation function of the k-gap field in the $k$-th term of the series. We have
\begin{align}
\frac{\delta^{2}\mathbb{Q}\left[Z(h,j)^{k}\right]
}{\delta j^{(k)}_{\psi}(t,\mathbf{x})\delta j^{(k)}_{\psi}(t',\mathbf{x}')}\Bigg|_{j^{(k)}_{\psi}=0}=iG_{\psi^{(k)}}^{(2)}(t,\mathbf{x};t',\mathbf{x}';k).
\label{23aa}
\end{align}
Also, $G^{(2)}_{\phi^{(k)}_l}(t,\mathbf{x};t',\bold{x}')$ is defined as the causal correlation function for the $l$-th phononic field in the $k$-th term of the series. It is written as
\begin{align}
\frac{\delta^{2}
\mathbb{Q}\left[Z(h,j)^{k}\right]}{\delta j^{(k)}_{\phi_l}(t,\mathbf{x})\delta j^{(k)}_{\phi_l}(t',\mathbf{x}')}\Bigg|_{j^{(k)}_{\phi_l}=0}=iG_{\phi^{(k)}_l}^{(2)}(t,\mathbf{x};t',\mathbf{x}').
\label{23aaa}
\end{align}  
A straightforward calculation gives that the causal two-point correlation function of the k-gap field in the $k$-th term of the series is written as 

\small\begin{eqnarray}
&&G^{(2)}_{\psi^{(k)}}(t,\mathbf{x};t',\mathbf{x}';k)=
\mathcal{N}_{0}(k)\int\mathcal{D}\psi^{(k)}~\psi^{(k)}(t,\mathbf{x})~\psi^{(k)}(t',\mathbf{x}')\nonumber\\
&&\times\exp\left[\frac{i}{2}\int d^{4}x~\psi^{(k)}(t,\mathbf{x})\left(\Box-k\sigma^{2}\right)\psi^{(k)}(t,\mathbf{x})\right].
\label{34}
\end{eqnarray}
\normalsize In the same way, it is possible to show that the causal two-point correlation function of the $l-$th phononic field in the $k$-th term of the series is written as  $G_{\phi^{(k)}_l}^{(2)}(t,\mathbf{x};t',\mathbf{x}')
$ is
\begin{align}
G_{\phi^{(k)}_l}^{(2)}(t,\mathbf{x};t',\mathbf{x}')&=\mathcal{N}^{(k)}_{l}
\int\mathcal{D}\phi^{(k)}_l~\phi^{(k)}_l(t,\mathbf{x})\phi^{(k)}_l(t',\mathbf{x}')\nonumber\\
&\times \exp\left[\frac{i}{2}\int d^{4}x ~\phi^{(k)}_l(t,\mathbf{x})~\Box~\phi^{(k)}_l(t,\mathbf{x})\right].
\label{34n}
\end{align}
From now on we will assume that the multiplet of $k-1$ phononic fields $\phi^{(k)}_l$ has all the same elements, i.e., we define $\phi^{(k)}_2(t,\mathbf{x}) = \phi^{(k)}_3(t,\mathbf{x}) = \dots = \phi^{(k)}_k (t,\mathbf{x})\equiv \phi^{(k)}(t,\mathbf{x})$. Therefore, $\mathcal{N}^{(k)}_{2} = \mathcal{N}^{(k)}_{3} = \dots \equiv \mathcal{N}_k^{(k)}$. And in this way we will have $k-1$ equations equal to Eq.\eqref{34n}.

Using the previous results, we can write the two-point correlation function associated with the k-gap field and phononic fields. We have
\begin{equation}
\bar{G}_{\psi}^{(2)}(t,\mathbf{x};t',\mathbf{x}')=
\sum_{k=1}^{N}c_{k}^{(1)}G^{(2)}_{\psi^{(k)}}(t,\mathbf{x};t',\mathbf{x}';k),
\label{con1}
\end{equation}
and
\begin{equation}
\bar{G}_{\phi}^{(2)}(t,\mathbf{x};t',\mathbf{x}')=
\sum_{k=2}^{\infty}c_{k}^{(1)}\prod_{l=2}^{k}G^{(2)}_{\phi_{l}^{(k)}}(t,\mathbf{x};t',\mathbf{x}').
\label{con2}
\end{equation}
Note that in the Eq.\eqref{con1} the summation ends in $N$. This will be clarified in the next section. 

Since any field theory is determined by its correlation functions, the effects of the noise field are to produce free phononic and gapped momentum excitations in the liquid. There is an important point that we would like to stress. To discuss the regime $\omega\tau_{c}\approx 1$, we introduced a noise field which was integrated out. The k-gap field must describe the behavior of the collective modes in the kinetic regime.

\section{Specific heat in liquids using functional determinants}\label{sec: specific heat}

In \ref{sec:casimir}, we perform the canonical quantization of the phononic fields and also the k-gap fields defined by the functional series. These k-gap fields define gapped momentum states in our model. From Eq.\eqref{32}, we have two kinds of frequencies. The real frequencies, for $\bold{p}^{2}\geq k\sigma^{2}$, where $\omega_{k}(\bold{p})=(\bold{p}^{2}-k\sigma^{2})^{\frac{1}{2}}$, and  the imaginary frequencies for $\bold{p}^{2}<k\sigma^{2}$, where $\omega_{k}(\bold{p})=\mp i(k\sigma^{2}-\bold{p}^{2})^{\frac{1}{2}}$. For each term of the series, the frequencies can be written as 
\begin{align}
\omega_{k}(\mathbf{p})=&\mp iu~\theta(k\sigma^{2}-\bold{p}^{2})(k\sigma^{2}-\bold{p}^{2})^{\frac{1}{2}}\nonumber\\
&+u~\theta(\bold{p}^{2}-k\sigma^{2})\bigl(\bold{p}^{2}-k\sigma^{2}\bigr)^{\frac{1}{2}}.
\label{44a}
\end{align} 
For $k\sigma^{2}>\bold{p}^{2}$, the lifetime of the quasi-particles is dominated by decaying pole terms of the excitations with gaps in the pseudo-momentum space. We restrict ourselves to field operators possessing non-vanishing Fourier components only for $\bold{p}^{2}\geq k\sigma^{2}$. In this case, the energy spectrum of the model is real. This procedure eliminates the dispersive  contribution, and we are discussing only the collective non-decaying excitations. Each term in the series contains one field with dispersion relation that exhibits a gap in pseudo-momentum space. One defines the angular frequencies $\nu_{k}(\mathbf{p})$ associated with the gapped momentum excitations as 
\begin{equation}
\nu_{k}(\mathbf{p})=u~\left(\bold{p}^{2}-k\sigma^{2}\right)^{\frac{1}{2}}.
\label{44correct}
\end{equation}

Since the Debye work \cite{debye}, phonons have been connected to the thermodynamic properties of solids. One can ask whether the same approach developed for solids using phonons can be used to find the specific heat of liquids. First, we derive the temperature dependence of the specific heat in low-temperature liquids due to the contributions of phononic and k-gap excitations. To proceed, instead of hyperbolic operators, we are discussing elliptic operators in a compact domain.

We consider the liquid in the solid-like regime as a system of phonons and k-gap excitations. Hence, the heat capacity at constant volume $C_V$ can be generally expressed in terms of the averages of the partition function $Z(h)$ as follows:
\begin{equation}\label{CV first}
C_V=\beta^2\sum_{k=1}^{\infty}\frac{(-1)^{k+1}}{k! k}\frac{\partial^2}{\partial\beta^2}\mathbb{E}\left[Z(h)^{k}\right],
\end{equation}
where $\beta = 1/T$. The Euclidean actions for the phonon fields and the k-gap fields are Gaussian. The functional integrals yield functional determinants. The terms $\mathbb{E}\left[Z(h)^{k}\right]$ are given by the ratio of the regularized determinants of the corresponding operators:
\begin{equation}\label{eq:partition_function_k}
\mathbb{E}[Z(h)^{k}] =\frac{\left\{\mathrm{det}\left[-\frac{1}{\mu^2}\left( \frac{\partial^2}{\partial \tau^2} + \Delta\right)\right]\right\}^{\frac{1-k}{2}}}{\left\{\mathrm{det}\left[-\frac{1}{\mu^2}\left(\frac{\partial^2}{\partial \tau^2}+\Delta +k\sigma^2\right)\right]\right\}^{\frac{1}{2}}}, 
\end{equation}
where $\mu$ is a constant with the inverse dimension of length.

To proceed, the regularization of functional determinants is standard. If $\lambda_k$ is a sequence of non-zero real numbers, the zeta-regularized product of these numbers $\prod_k \lambda_k$ is given by:
\begin{equation}
    \prod_{k=0}^{\infty} \lambda_k \equiv e^{-\zeta'(0)}
\end{equation}
where
\begin{equation}
    \zeta(s) = \sum_{k=0}^{\infty} \lambda_k^{-s}, ~\text{Re}(s) > s_0.
\end{equation}
It is possible to analytically continue the spectral zeta-function to a meromorphic function in the whole complex plane. This allows one to give a meaning to the spectral zeta-function even outside the domain of convergence of the series. In particular, $s=0$ turns out to be a regular point such that the first derivative $\zeta'(0)$ is well defined. However, if $\lambda_k$ is the sequence of positive eigenvalues of the Laplacian on a smooth, bounded, open set $\Omega \subset \mathbb{R}^d$, then the regularized product is the determinant of the Laplacian \cite{seeley1, hawking77, quine93}. The spectral zeta-function related to the operator $-\frac{1}{\mu^2}\left(\frac{\partial^2}{\partial \tau^2} +\Delta\right)$,  is given by:
\begin{equation}\label{spectralz}
    \zeta(s) = \frac{V}{2\pi^2} \int_0^{p_D} dp ~p^2 \sum_{n\in \mathbb{Z}}\left[ \left(\frac{2\pi n}{\beta}\right)^2 + p^2\right]^{-s}.
\end{equation}
Also, the spectral zeta-functions related to operators $-\frac{1}{\mu^2}\left[\frac{\partial^2}{\partial \tau^2}+(\Delta -k\sigma^2)\right]$ with $k=1,2,\dots$ are
\begin{equation}\label{b1}
    \zeta_{k}(s)= \frac{V}{2\pi^2}\int_{\sigma \sqrt{k}}^{p_D} dp ~p^2\sum_{n\in \mathbb{Z}}\left[ \left(\frac{2\pi n}{\beta}\right)^2 + p^2 - k \sigma^2 \right]^{-s},
\end{equation}
where $|\mathbf{p}| = p$.

\begin{figure}[h!]
    \centering\includegraphics[width=1\linewidth]{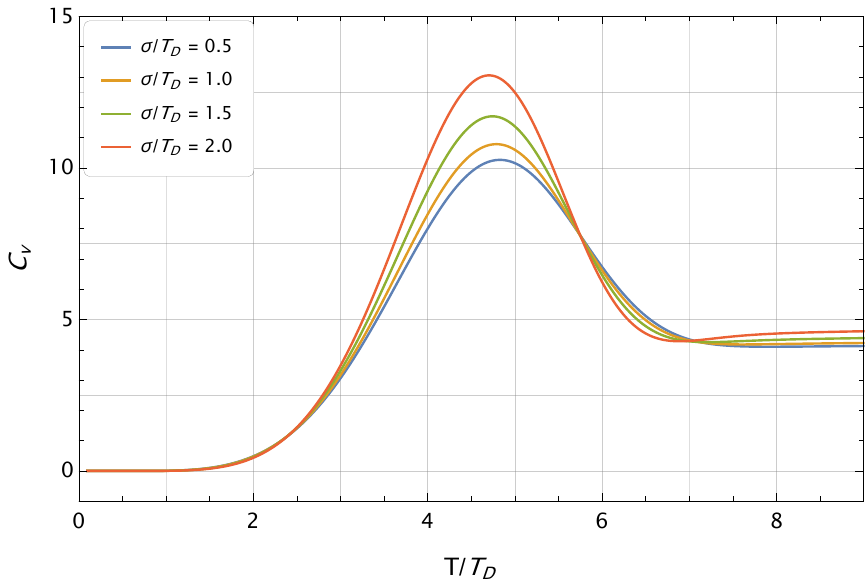}
    \caption{Plot of heat capacity, $C_V$, as a function of normalized temperature $T/T_D$ for different values of $\sigma/T_D$. Here we consider $V T^3_D = 1$ and ${\epsilon_2}/T_D = 0.001$. We assume $\epsilon_2$ is a regularization factor introduced in the calculation of the heat capacity and $T_D$ is the Debye temperature of the liquid. To generate the curves, we consider a summation of $k$ from 1 to 100.}
    \label{heat capacity}
\end{figure}

In \ref{sec: zeta regularization}, the regularization of the determinants given by Eq. \eqref{eq:partition_function_k} is discussed in more detail. All the mathematical tools needed to answer the question of how the non-hydrodynamic degrees of freedom contribute to the thermodynamics of liquids are now assembled. In our approach, the quasi-particle excitations in the fluid give the heat capacity of the liquids, which can be seen in Fig. \ref{heat capacity}.

Liquids with different structures exhibit different heat capacities as a function of temperature. The phonon model has been used to describe simple liquids whose heat capacity decreases with temperature. In Ref. \cite{Liu2}, the heat capacities of $Ar$, $N_2$, $Na$ and $CH_4$ were presented.  Finally, our analysis reveals the role of k-gap fields in defining the behavior of the specific heat of simple liquids as a function of temperature.

\section{Conclusions and perspectives}\label{sec:conclusions}

As we discussed, in spite of considerable efforts, a unifying physical model of thermodynamic properties of liquids is still under construction, due to the complexity of liquid behavior at different scales, such as, for example, viscoelastic properties. There is an interpolation between the purely elastic solid behavior and the purely dissipative response of a fluid. Maxwell discussed a simple model for viscoelastic materials, which is characteristic of liquids. There is a viscous flow on long time scales and elastic behavior on very short time scales. To discuss a solid-like behavior in liquids, Frenkel defines $\tau_{f}$, the liquid relaxation time, i.e., the average time that atoms/molecules spend traversing the interatomic/intermolecular spacing. For times shorter than $\tau_{f}$, the behavior of the system is that of a disordered solid with shear elastic waves. This construction led to the theory of gapped momentum states in liquids. It is important to point out that other authors long time ago suggested that a liquid resembled a disordered crystal \cite{marc,casals}.

In the present work, we established, in low-temperature liquids, the connection between gapped momentum states of elementary excitations and hydrodynamics as an effective field theory. To achieve such a connection, we use the functional integral formalism of field theory. Here, we aim to model the behavior of the elementary excitations of the liquid taking into account processes for frequencies satisfying $\omega\tau_{c}\geq 1$, i.e., when non-hydrodynamic degrees of freedom are dominant. To model such degrees of freedom, we are using an additive, delta-correlated (in space and time) noise field. We use the fact that liquids at small scales or short time scales must be modeled as dynamically disordered solids. If we are interested in describing fluids at the intramolecular scale, one has to introduce a colored noise.

Our functional approach to emergent phononic and tachyonic-like excitations in liquids can be viewed as complementary to recent symmetry-based interpretations. Notably, Baggioli et al. \cite{baggioli2022} have developed a unified topological field theory that explains the appearance of the $k$-gap in liquids through the phase relaxation of Goldstone modes. Their work reveals that the fundamental distinction between solids and liquids lies in the conservation (or lack thereof) of a two-form current related to the single-valuedness of the displacement field. This perspective aligns with our findings on the emergence of gapped momentum states due to quantum fluctuations. Our theoretical framework predicts the low-temperature thermodynamic signature of k-gap fluids, i.e., the temperature dependence of the specific heat. This behavior emerges naturally from the distributional zeta-function approach when analyzing how excitations with gaps in the pseudo-momentum states contribute to thermodynamic properties. Unlike crystalline solids with their characteristic Debye $T^3$ law arising from three acoustic phonon branches, liquids with $k$-gapped transverse modes effectively possess a reduced vibrational density of states at low frequencies. Although propagating transverse modes with these new dispersion relations have not yet been experimentally detected in classical liquids, there are indirect pieces of evidence supporting that the tachyonic transverse modes do exist. Nevertheless, these modes were reported in dusty plasmas, granular fluids, and colloidal systems \cite{nosenko, jiang, bai}. The thermodynamics of solids has been developed using phonons. By incorporating the effects of quantum noise on hydrodynamic fields, our model captures the essential physics that the phase space available to phonons in liquids is not fixed but variable. The emergence of these new excitations with gaps in the pseudo-momentum space through noise-induced processes thus offers a concrete mechanism for the temperature-dependent changes in the available vibrational modes.

A natural continuation of this work is to discuss the spectral density of fluids due the phononic and k-gap excitations, as discussed in Refs. \cite{zaccone21, stamper, brazhkin24}. 

\section*{Acknowledgements}
The authors are grateful to G. O. Heynmans, G. Scorza, G. C. D. Rodrigues-Camargo, C. Farina, M. M. Balbino, B. F. Svaiter and G. Krein for fruitful discussions. This work was partially supported by Conselho Nacional de Desenvolvimento Cient\'{\i}fico e Tecnol\'{o}gico - CNPq, 303436/2015-8 (NFS), 307626/2022-9 (AMSM) and FAPERJ PhD merit fellowship - FAPERJ Nota 10, 203.709/2025 (FS). The author I. P. de Freitas are also thankful for the financial support of CNPq.

%% If you have bibdatabase file and want bibtex to generate the
%% bibitems, please use
%%

%% The Appendices part is started with the command \appendix;
%% appendix sections are then done as normal sections
\appendix

\section{The canonical quantization of the hydrodynamic fields and fields with gaps in the pseudo-momentum space}\label{sec:casimir}

As presented in Sec.\ref{sec: specific heat}, we will perform the canonical quantization of the phononic fields and also the k-gap fields defined by functional series.

Comparing  Eq.\eqref{frenkel} with Eq.\eqref{44correct} shows that the behavior of the collective excitations resembles the gapped momentum states in the solid-like elastic regime of wave propagation.
The result from the literature, where the volume of the phase space available to collective excitations reduces with temperature, is here related to the strength of the noise. 
One important point is that there is a threshold for the pseudo-momentum $|\bold{p}_{c}|$ where, above such a critical value, there is a breakdown of the linear dispersion relation, i.e., $|\bold{p}_{c}|\leq\frac{2\pi}{l_{c}}$, where $l_{c}$ is the mean free path of the constituents of the liquid.  This critical pseudo-momentum defines a critical $k$ in the functional series that we call $N$ in the quantized hydrodynamic model. We have 
\begin{equation}
N=k_{c}=\lfloor\, \sigma^{-2}|\bold{p}_{c}|^{2}\rfloor,
\label{45}
\end{equation}
where $\lfloor\xi\rfloor$ denotes the largest integer $\leqq \xi$. The positive frequency modes associated with the k-gap fields are $v^{(k)}_{\bold{p}}(t,\mathbf{x})$, where
\begin{equation}
v^{(k)}_{\bold{p}}(t,\mathbf{x}) = e^{i\big(\bold{p}.\mathbf{x} -\nu_{k}(\bold{p}) t\big)},
\label{48a}
\end{equation}
The positive frequency modes associated with the k-gap fields are $v^{(k)}_{\lambda\bold{p}}(t,\mathbf{x})$, where
\begin{equation}
v^{(k)}_{\lambda\bold{p}}(t,\mathbf{x}) = e^{i\big(\bold{p}.\mathbf{x} -\nu_{k}(\bold{p}) t\big)}.
\label{48b}
\end{equation}
As defined in Eq.\eqref{eq: l and t modes}, the sound wave has longitudinal and transverse components. We were studying only the transverse components of the wave $u_i(t,\mathbf{x})$. By introducing an index $i$ and the polarization vectors $e^{(\lambda)}_i$ in our k-gap fields, we can connect it to our initial quantity $u_i(t,\mathbf{x})$. The $\lambda=2,3$ denote the transverse polarizations, and since the noise has only transverse components, the tachyonic field has the same structure.

The Fourier expansion of the k-gap field operator $\psi^{(k)}_{i}(t,\mathbf{x})$ (for $i=2,3$) is given by
\small\begin{align}
&\psi^{(k)}_{i}(t,\mathbf{x})\nonumber\\
&=\sum_{
\bold{p}^{2}\geq k\sigma^{2}}\sum_{\lambda=2}^{3}i ~\left(\frac{\nu_{k}(\mathbf{p})}{2V}\right)^{\frac{1}{2}}e_{i}^{(\lambda)}\nonumber\\
&\times \left(b^{(k)}_{\lambda\bold{p}}v^{(k)}_{\lambda\bold{p}}(t,\mathbf{x}) 
-b^{(k)\dagger}_{\lambda\bold{p}}
v^{(k)*}_{\lambda\bold{p}}(t,\mathbf{x})\right),\label{49}
\end{align}
\normalsize where $b^{(k)}_{\lambda\bold{p}}$ and  $\bigl(b^{(k)}_{\lambda\bold{p}}\bigr)^{\dagger}$ are the annihilation and creation operators for the k-gap fields, with polarization $\lambda$. As we discussed before, the $e^{(\lambda)}_{i}$ are the transverse polarization vectors, where $p^{i}e_{i}^{(\lambda)}=0$, for $\lambda=2,3$ and $p^{i}e_{i}^{(\lambda)}=|\bold{p}|$, for $\lambda=1$.  We have $e^{(1)}_{i}=(1,0,0)$, $e^{(2)}_{i}=(0,1,0)$ and $e^{(3)}_{i}=,0,1)$. 
The commutation relation between the annihilation and creation operators are the standard ones given by
\begin{equation}
\bigl[b^{(k)}_{\lambda\bold{p}},b^{(k)}_{\lambda'\bold{p}'}\bigr]=0,\,\,\,\,\bigl[b^{(k)\dagger}_{\lambda\bold{p}},b^{(k)\dagger}_{\lambda'\bold{p}'}\bigr]=0,
\end{equation}
and
\begin{equation}
\bigl[b^{(k)}_{\lambda\bold{p}},b^{(k)\dagger}_{\lambda'\bold{p}'}\bigr]=\delta_{\lambda\lambda'}\delta(\bold{p}-\bold{p}').
\end{equation} 
Therefore we get new kinds of vacuum states. The old one $|\Omega_{0}\rangle$ and the ones $|\Omega^{(k)}_{b}\rangle$ associated for each term of the functional series defined by
\begin{equation}
b^{(k)}_{\lambda\bold{p}}~|~\Omega^{(k)}_{b}\rangle=0,  ~\forall ~\bold{p}.
\end{equation} 
Using the functional series, the k-gap Fock vacuum state can be written, making use of vacuum Fock states of each term of the series. In this case, there is an upper bound in the absolute values of pseudo-momentum space. We have the vacuum states
\begin{equation}
|~\Omega^{(1)}_{b}\rangle,~|~\Omega^{(2)}_{b}\rangle,~\dots,~|~\Omega^{(N-1)}_{b}\rangle ~|~\Omega^{(N)}_{b}\rangle.
\end{equation}
We write
\begin{equation}
|~\Omega_{T}\rangle=|~\Omega^{(1)}_{b}\rangle\otimes|~\Omega^{(2)}_{b}\rangle\otimes\dots\otimes|~\Omega^{(N-1)}_{b}\rangle\otimes|~\Omega^{(N)}_{b}\rangle.
\end{equation}

The Hamiltonian of the k-gap fields defined by the functional series is given by
\begin{align}
H_{T}=\frac{1}{2}\sum_{k=1}^{N}\sum_{\bold{p}^{2}\geq k\sigma^{2}}
\sum_{\lambda=2}^{3}\nu_{k}(\bold{p})~\biggl(b^{(k)\dagger}_{\lambda\bold{p}}
b^{(k)}_{\lambda\bold{p}}+b^{(k)}_{\lambda\bold{p}} b^{(k)\dagger}_{\lambda\bold{p}}\biggr),
\label{53}
\end{align}
where we are using Eq.\eqref{45}. Since the ground states of the new field sector of the theory are cyclic with respect to the polynomials of the fields, an arbitrary state of the Hilbert space is constructed with the generic $|\Omega^{(k)}_{b}\rangle$ Fock vacuum state and is given by a linear superposition of multi-elementary excitations with gaps in the pseudo-momentum space, which we denote as $|\aleph_\lambda^{(k)}\rangle$
\small\begin{align}
&|~\aleph_\lambda^{(k)}\rangle\nonumber\\
&=\sum_{q=0}^{\infty}\frac{1}{\sqrt{q!}}\int\vartheta^{(k)}_{q}({\bold{p}}_{1},\dots,{\bold{p}}_{q})\,b^{(k)\dagger}_{\lambda\bold{p}_{1}}\dots b^{(k)\dagger}_{\lambda\bold{p}_{q}}\prod_{i=1}^{q}(d^{3}p_{i})|~\Omega_{b}^{(k)}\rangle,
\end{align}
\normalsize where $\vartheta_{0}^{(k)} \in \mathbb{C}$ and $\vartheta_{q}^{(k)}$ for $q\geq 2$ are symmetric functions. We have
\small\begin{equation}
\langle \aleph_\lambda^{(k)}~|~\aleph_\lambda'^{(k)}\rangle=\delta_{\lambda\lambda'}\sum_{q=0}^{\infty}\int|\vartheta^{(k)}_{q}({\bold{p}}_{1},\dots,{\bold{p}}_{q})|^{2}\prod_{i=1}^{q}(d^{3}p_{i})<\infty.
\end{equation} 
\normalsize The above representation defines the Fock space that can be generated using the Fock vacuum state $|~\Omega_{b}^{(k)}\rangle$.
Note that due to the condition $|\bold{p}|\geq k\sigma^{2}$, the set of functions that we used to expand the field operator does not form a complete set. We have to impose the new completeness relation
\begin{equation}
\sum_{\bold{p}^{2}\geq k\sigma^{2}}
v^{(k)}_{\lambda\bold{p}}(t,\mathbf{x}) 
v^{(k)*}_{\lambda\bold{p}}(t',\mathbf{x}')
\Big|_{t=t'=0}=f_{k}(\bold{x}-\bold{x'})
\end{equation}
where
\begin{equation}
f_{k}(\bold{x}-\bold{x}')=\delta^{3}(\bold{x}-\bold{x}')+\frac{1}{2\pi^{2}}
\frac{k\sigma^{2}\cos k\sigma^{2}-\sin k\sigma^{2}}{k\sigma^{2}|\bold{x}-\bold{x}'|}.
\end{equation}

With the new completeness relation one can  calculate the expectation value of the Dyson time-ordered product associated with the tachyon-like field in the $k-$th term of the functional series. 
\begin{equation}
G^{(2)}_{\psi_{ij}^{(k)}}(t,\mathbf{x};t',\bold{x}')=-i\langle \Omega_{b}~|T[\psi_{i}^{(k)}(t,\mathbf{x}) \psi_{j}^{(k)}(t',\mathbf{x}')]|~\Omega_{b}\rangle\delta_{ij}.
\label{nova}
\end{equation}
Substituting the Fourier representation of the k-gap field operator, defined in Eq.\eqref{49}, in the Eq.\eqref{nova} one obtains the causal propagator for the field in the $k$-th term of the series  in this model. It is written as 
\small\begin{align}
&G^{(2)}_{\psi^{(k)}}(t,\mathbf{x};0,\bold{0})\nonumber\\
&=-\frac{i}{V}\sum_{
\bold{p}^{2}\geq k\sigma^{2}}
\sum_{\lambda=2}^{3}\frac{\nu_{k}(\mathbf{p})}{2}~\Bigl(\theta(t)
v^{(k)}_{\lambda\bold{p}}(t,\mathbf{x}) +\theta(-t)v^{(k)*}_{\lambda\bold{p}}(t,\mathbf{x})\Bigr),\label{58}
\end{align}

\normalsize One can write the Fourier components of the above equation. 
We can write the Fourier representation of the two-point correlation function of the model with the contribution coming from all k-gap fields represented by $G_{\psi^{(k)}}^{(2)}(\upsilon_\lambda,\bold{p})$.  
We have
\begin{eqnarray}
&&\bar{G}^{(2)}_{\psi}(\upsilon,\bold{p})
=\nonumber\\
&&\frac{1}{2}\sum_{k=1}^{N}c_{k}^{(1)}\nu_{k}(\bold{p})~\left(\frac{1}{\upsilon-\nu_{k}(\bold{p})+i\delta}-
\frac{1}{\upsilon_\lambda+\nu_{k}(\bold{p})-i\delta}\right), \nonumber\\
\end{eqnarray}
where again the infinitesimal term in the denominator indicates in what half-plane of complex frequency the corresponding integrals will converge.
We rewrite the above equation in terms of sums over $\bold{p}$, where the volume $V$ appears. We have
\begin{equation}
\bar{G}^{(2)}_{\psi}(\upsilon)=\frac{1}{V}\sum_{k=1}^{N}c_{k}^{(1)}
\sum_{
\bold{p}^{2}\geq k\sigma^{2}}\frac{\nu_{k}^{2}(\bold{p})}{\upsilon^{2}-\nu_{k}^{2}(\bold{p})+i\delta}.
\end{equation}
From the dispersion relation of $\nu_{k}(\bold{p})$ given by Eq.\eqref{44correct} for large $k$'s the contribution of k-gap fields vanishes, due to the breakdown of the linear dispersion relation for large pseudo-momentum. Therefore in the summation we have a finite number of terms. 
The discrete sum over $\mathbf{p}$ can be replaced by a continuous integral using that
\begin{equation}
\sum_{\mathbf{p}}\rightarrow \frac{V}{(2\pi)^{3}}\int d^{3}q.
\end{equation}
Introducing a Debye quasi-momentum cut-off  $q_{D}$ we write $\bar{G}^{(2)}_{\psi}(\upsilon)$ as 
\begin{align}
\bar{G}^{(2)}_{\psi}(\upsilon)=\frac{u^{2}}{2\pi^{2}}
\sum_{k=1}^{N}c_{k}^{(1)}
\int_{k\sigma^{2}}^{q_{D}}dq \frac{q^{4}-k\sigma^{2}q^{2}}{\upsilon^{2}+u^2(k\sigma^{2}-q^{2})}
.
\end{align} 
It is clear that the same discussion can be performed for the phononic field.  We can write the Fourier representation for the causal correlation functions.  The total contribution of the phonons in the functional series can be written as  
\begin{equation}
\bar{G}_{\phi}^{(2)}(\upsilon,\bold{p})=
\sum_{k=2}^{\infty}c_{k}^{(1)}\Biggl[\frac{\omega^{2}(\bold{p})}{\upsilon^{2}-\omega^{2}(\bold{p})+i\delta}\Biggr]^{k-1}.
\label{43}
\end{equation}
The above expression is quite interesting.
Again, it is possible to rewrite the above equation in terms of sums over all the pseudo-momentum $\bold{p}$. We get 
\begin{align}
\bar{G}_{\phi}^{(2)}(\upsilon)= \frac{1}{V}
\sum_{\bold{p}}
\sum_{k=2}^{\infty}c_{k}^{(1)}\biggl(\frac{\omega^{2}(\bold{p})}{\upsilon^{2}-\omega^{2}(\bold{p})+i\delta}\biggr)^{k-1}.
\label{433}
\end{align}

\section{Euclidean actions for phonons and fields with gaps in the pseudo-momentum space and its regularized determinants}\label{sec: zeta regularization} 

As presented in Sec.\ref{sec: specific heat}, we will resort to the distributional zeta-function method to investigate the influence of the k-gap momentum states in the specific heat capacity. In Eq.\eqref{eq:partition_function_k}, there are two different operators.  First, the operator $A \equiv \left[-\frac{1}{\mu^2}\left(\frac{\partial^2}{\partial \tau^2} + \Delta\right)\right]$, for which the spectral  zeta-function related to this operator is given by:
\begin{equation}\label{spectralzCorrect}
    \zeta(s) = \frac{V}{(2\pi)^3} \int_0^{\mathbf{p}_D} d\mathbf{p} \sum_{n\in \mathbb{Z}}\left[ \left(\frac{2\pi n}{\beta}\right)^2 + \mathbf{p}^2\right]^{-s}.
\end{equation}

From now on, we will use $\beta = 1/T$. Using Eq.\eqref{spectralzCorrect}, one can rewrite the spectral zeta-function $\zeta(s)$ as:
\begin{equation}\label{zeta1}
    \zeta(s) = \frac{V}{\pi^2} \left(2\pi T\right)^{3-2s}\int_0^{\frac{T_D}{T}} dx \, x^2 \sum_{n \in \mathbb{Z}}\left(n^2+x^2\right)^{-s},
\end{equation}

\noindent where $T_D$ is assumed to be the Debye temperature for the liquid. The summation $\sum_{n \in \mathbb{Z}}(n^2+x^2)^{-s}$ is an Epstein-Hurwitz zeta-function. Its analytic continuation is given by \cite{fordnami}:
\begin{align}\label{zetahw}
    \sum_{n \in \mathbb{Z}}(n^2+x^2)^{-s} = x^{1-2s} &\left[\sqrt{\pi}\frac{\Gamma(s-1/2)}{\Gamma(s)} \right. \nonumber\\
    &+ \left. 4\sin(\pi s)\int_1^\infty dt\, \frac{\left(t^2-1\right)^{-s}}{e^{2\pi x t}-1}\right].
\end{align}

With the eigenvalues of the operator $A$, one can define $\mathrm{det}A$, which is exactly the product of the eigenvalues. This product is regularized by the usual spectral zeta-function method.  
By applying Eq.\eqref{zetahw} to  \eqref{zeta1}, it is possible to verify that $\zeta(0)= 0$. Therefore, the determinant is scaling independent. The derivative $\zeta'(0)$ is:
\begin{equation}
    \zeta'(0) = 32\pi^2 VT^3\int_0^{\frac{T_D}{T}} dx \, x^3 \int_1^\infty \frac{1}{e^{2\pi xt}-1}
\end{equation}

In order to solve the above integration, we can introduce a minimum cut-off $\epsilon_1$ in the integration over $x$. This procedure is useful since the above integral is undetermined at the limits of integration. Therefore, the resulting integral is:
\small \begin{eqnarray}
    &&\zeta'(0) = -16\pi VT^3 \int_{\epsilon_1}^{\frac{T_D}{T}} dx \, x^2 \log{(1-e^{-2\pi x})}\nonumber\\
    && -\frac{16\pi}{3} VT^3 \left\{\left[x^3\log{\left(1-e^{-2\pi x}\right)}\right]_{\epsilon_1}^{\frac{T_D}{T}} -\int_{\epsilon_1}^{\frac{T_D}{T}} dx \, \frac{1}{e^{2\pi x}-1}\right\} \nonumber\\
\end{eqnarray}

\normalsize Considering $\epsilon_1 \ll 1$, we have:
\begin{equation}\label{eq: zeta prime A}
    \frac{\zeta'(0)}{V T^3} =\frac{32\pi^2}{3}\int_0^{\frac{T_D}{T}} dx \, \frac{x^3}{e^{2\pi x}-1}- \frac{16\pi}{3}\log{\left(1-e^{-\frac{2\pi T_D}{T}}\right)}.
\end{equation}

Therefore, for $A \equiv \left[-\frac{1}{\mu^2}\left(\frac{\partial^2}{\partial \tau^2} + \Delta\right)\right]$, the regularized value of $\mathrm{det} A $ is: 
\begin{equation}\label{eq: detA}
    \mathrm{det}A = \exp \zeta'(0), 
\end{equation}
where $\zeta'(0)$ is given by Eq.\eqref{eq: zeta prime A}.

The operator $B\equiv \left[-\frac{1}{\mu^2}\left(\frac{\partial^2}{\partial \tau^2}+\Delta -k\sigma^2\right)\right]$ is related to the following spectral zeta-function
\begin{equation}\label{b1correct}
    \zeta_{k}(s) = \frac{4\pi V}{(2\pi)^3} \int_{\sigma \sqrt{k}}^{p_c} dp \,p^2\,\sum_{n\in \mathbb{Z}}\left[ \left(2\pi n T\right)^2 + p^2 - k \sigma^2 \right]^{-s},\nonumber\\
\end{equation}
where $|\mathbf{p}| = p$.

The procedure to solve Eq.\eqref{b1correct} is very similar to the one presented previously. However, in order to avoid infrared divergences, it is suitable to choose a minimum cutoff $\epsilon_2$. Therefore, by proceeding in an analogous way to rewriting Eq.\eqref{b1},
\begin{align}
    \zeta_k(s)
    =& \frac{V}{\pi^2} \left(2\pi T\right)^{3-2s} \int_{\epsilon_2}^{\frac{T_D}{T}} dx \, x^{2-2s}\sqrt{x^2 +k\left(\frac{ \sigma}{2\pi T}\right)^2}\nonumber\\
    &\times \left[\pi\frac{\Gamma(s-1/2)}{\Gamma(s)} + 4\sin(\pi s)\int_1^\infty dt \frac{1}{e^{2\pi xt} -1}\right].\nonumber\\
\end{align}

With the eigenvalues of the operator $B$, one can define $\mathrm{det}B$, which is exactly the product of the eigenvalues. This product is regularized by the usual spectral zeta-function method. It is possible to verify that $\zeta_k(0)=0$, and $\zeta_k'(0)$ is given by:
\begin{eqnarray}
    \zeta_k'(0) &=& 32\pi^2 VT^3\int_{\epsilon_2}^{\frac{T_D}{T}} dx \, x\sqrt{x^2 + k\left(\frac{\sigma}{2\pi T}\right)^2} \nonumber\\
    &&\times \int_1^\infty \frac{1}{e^{2\pi xt}-1}
\end{eqnarray}

\noindent Finally, defining $f(y) = \log{\left(1- e^{-2\pi \,y}\right)}$, the result for $\zeta_k'(0)$ is
\begin{align}
    &\frac{\zeta_k'(0)}{VT^3} \nonumber\\
    &= \frac{16\pi}{3} \Bigg\{f(\epsilon_2)\left[{\epsilon_2^2}+k\left(\frac{\sigma }{2\pi T}\right)^2\right]^{\frac{3}{2}}  \nonumber\\
    &- f\left(\frac{T_D}{T}\right) \left[\frac{T_D^2}{T^2}+k\left(\frac{\sigma }{2\pi T}\right)^2\right]^{\frac{3}{2}} \nonumber\\
    & \left.+2\pi\int_{\epsilon_2}^{\frac{T_D}{T}} dx \, \left[x^2+k\left(\frac{\sigma }{2\pi T}\right)^2\right]^{\frac{3}{2}} \frac{1}{e^{2\pi x}-1}\right\} \nonumber\\
    \label{eq: zeta prime B}
\end{align}

For $B \equiv \left[-\frac{1}{\mu^2}\left(\frac{\partial^2}{\partial \tau^2} + \Delta-k \sigma^2\right)\right]$, the regularized value of $\mathrm{det} B $ is: :
\begin{equation}\label{eq: detB}
    \mathrm{det}B = \exp \zeta'_k(0), 
\end{equation}
where Eq.\eqref{eq: zeta prime B}.

 The expression for $\mathbb{E}[Z^k(h)]$, given by Eq.\eqref{eq:partition_function_k}, can be constructed using Eqs.\eqref{eq: detA} and \eqref{eq: detB}. I.e.,:
\begin{equation}
    \mathbb{E}[Z^k(h)] = \exp\left\{\frac{1}{2} \Big[ (1-k)\zeta'(0) - \zeta'_k(0) \Big] \right\},
\end{equation}
where $\zeta'(0)$ and $\zeta'_k(0)$ are given, respectively, by Eqs.\eqref{eq: zeta prime A} and \eqref{eq: zeta prime B}.

%% else use the following coding to input the bibitems directly in the
%% TeX file.

%%\begin{thebibliography}{00}

%% \bibitem[Author(year)]{label}
%% For example:

%% \bibitem[Aladro et al.(2015)]{Aladro15} Aladro, R., Martín, S., Riquelme, D., et al. 2015, \aas, 579, A101

%%\end{thebibliography}

\end{document}